# How perception of being recognized or not recognized by instructors as a "physics person" impacts male and female students' self-efficacy and performance

Yangqiuting Li, Kyle Whitcomb, and Chandralekha Singh

*Department of Physics and Astronomy University of Pittsburgh, Pittsburgh PA 15260*

**Abstract:** We discuss a study in a first year college introductory physics course for physical science and engineering majors that shows that women, on average, feel less recognized by their physics instructors than men as students who can excel in physics. We also discuss how this lack of perceived positive recognition pertaining to physics can adversely affect their self-efficacy and performance in the course. We recommend that physics instructors not be parsimonious in their praise of students and make a conscious effort to positively recognize their students for their effort and progress whenever an opportunity arises. Interviews with female students suggest that instructors should be careful not to give unintended messages to students, e.g., by praising some students for brilliance or intelligence as opposed to their effort because praising a student for brilliance can convey to other students that they do not have what is required to excel in physics. Interviews also suggest that when students ask instructors for help on physics problems, if instructors inadvertently label those problems as "easy", "trivial" or "obvious", it can also make students feel disparaged. The perception of being belittled by these kinds of unintended comments by instructors as well as a lack of positive recognition for good effort and progress have the potential to most adversely impact students from underrepresented groups including women.

**Introduction:** Physics has historically been portrayed as a field for brilliant men. Many prior studies have focused on the reasons for women's underrepresentation in physics and related disciplines from different perspectives and strategies to improve the learning environments so that all students can excel in physics courses [1-17]. Due to societal stereotypes, women often have significantly lower physics self-efficacy than men even when they perform similarly and many shy away from physics related majors and careers [5,6,9,13]. Moreover, being recognized by the instructor as a student who can excel in physics can be valuable for all students in physics courses. However, it is particularly important for underrepresented students including women and ethnic and racial minorities partly due to the societal stereotypes associated with who can excel in physics and lack of role models [1-24].

Perceptions regarding lack of positive recognition pertaining to whether a student can excel in physics and unintended belittling of students by instructors has a greater potential to negatively impact underrepresented students including women. For example, Eileen Pollock, the first woman to get a BS degree in physics at Yale University, decided to pursue graduate work in English despite finishing her physics undergraduate degree summa cum laude. In her memoir [25], she recounted the negative impact of not being positively recognized by her instructors, "By the start of my senior year, I was at the top of my class, with the most experience conducting research. But not a single professor asked me if I was going on to graduate school. When I mentioned shyly to Professor Zeller that my dream was to apply to Princeton and become a theoretician, he shook his head and said that if you went to Princeton, you had better put your ego in your back pocket, because those guys were so brilliant and competitive that you would get that ego crushed, which made me feel as if I weren't brilliant or competitive enough to apply." Being a woman, it is not surprising that Pollock would interpret such statements to imply that she was essentially being told that she was not capable of excelling like the brilliant men at Princeton!

Pollock [25] also noted lack of positive recognition from her thesis advisor, "Not even the math professor who supervised my senior thesis urged me to go on for a Ph.D. I had spent nine months missing parties, skipping dinners and losing sleep, trying to figure out why waves — of sound, of light, of anything — travel in a spherical shell, like the skin of a balloon, in any odd-dimensional space, but like a solid bowling ball in any space of even dimension. When at last I found the answer, I knocked triumphantly at my adviser's door. Yet I don't remember him praising me in any way. I was dying to ask if my ability to solve the problem meant that I was good enough to make it as a theoretical physicist. But I knew that if I needed to ask, I wasn't." She added that she was "certain this meant I wasn't talented enough to succeed in physics, I left the rough draft of my senior thesis outside my adviser's door and slunk away in shame." This example illustrates another missed opportunity in which a thesis advisor failed to positively recognize her accomplishments and she went from feeling "triumphant" about having solved her thesis problem to feeling she wasn't talented enough to succeed in physics. What is also worth noting is that while writing her book, when Pollock asked her former advisor what he thought of her thesis, he stated that it was exceptional and when pressed further confessed that he had never encouraged *anyone* to pursue further studies.

Here we discuss a study on how the perception of being positively recognized or not recognized appropriately by the instructor or teaching assistant (TA) as a "physics person" or a person who can excel in physics impacts male and female students' self-efficacy and performance at the end of a two-term college calculus-based introductory physics sequence. At the large public university where this study was conducted, Physics 1 involves mechanics and waves and physics 2 involves electricity and magnetism. A majority of students were first year engineering, physics, chemistry and math majors.

In addition to individual semi-structured think-aloud interviews with 30 student volunteers (20 women and 10 men), we used Structural Equation Modeling (SEM) with gender mediation. SEM is an approach involving simultaneous multiple regressions and can be used to predict relationships among different variables [26]. We focused on matched students who took both physics 1 and physics 2, i.e., there were 233 female and 464 male students followed from physics 1 to the end of physics 2. In four sections of the courses from which we present data, four instructors were involved who primarily used traditional lecture-based method. As shown in Fig. 1, we used this method [26] to investigate how male and female students' self-efficacy at the end of physics 2 (post self-efficacy 2 or Post SE 2) and their physics 2 grade (Grade 2) are mediated by the perceived recognition as a "physics person" by the course instructor/TA. Fig. 1 shows that our model controls for student high school math SAT scores (SAT-Math), high school GPA (HS-

GPA), physics 1 grade (Grade 1) and self-efficacy at the end of physics 1 (post self-efficacy 1 or Post SE 1). We collected data from a validated motivational survey [3, 12, 13, 18-20] to measure students' self-efficacy at the end of both physics 1 and 2 and students' perceived recognition by the instructor and TA at the end of physics 2. However, details of the survey and measurement part of SEM will be described elsewhere.

TABLE I. Mean values of high school GPA and SAT math scores, college grades and post self-efficacy in physics 1 and 2, and perceived recognition in physics 2 by gender, along with p-values showing statistical significance of *t*-tests and effect sizes (given by Cohen's d with a positive value favoring male students) showing the strength of gender contrast [26]. Score ranges for all variables are also shown.

| Predictors and Outcomes (Score Range) | Mean | | p value | Cohen's *d* |
|---|---|---|---|---|
| | Male | Female | | |
| High School GPA (0-5) | 4.20 | 4.34 | < 0.001 | -0.34 |
| SAT Math (400-800) | 713 | 706 | 0.130 | 0.13 |
| Post Self-Efficacy in Physics 1 (1-4) | 3.06 | 2.83 | < 0.001 | 0.47 |
| Physics 1 Grade (0-4) | 2.93 | 2.74 | 0.001 | 0.28 |
| Perceived Recognition (1-4) | 2.55 | 2.14 | < 0.001 | 0.55 |
| Post Self-Efficacy in Physics 2 (1-4) | 2.91 | 2.65 | < 0.001 | 0.47 |
| Physics 2 Grade (0-4) | 2.73 | 2.48 | < 0.001 | 0.31 |

**Results**: Table I shows the averages and effect sizes [26] for the differences between male and female students for their high school GPA and SAT math scores, their college physics 1 and physics 2 grades, their post self-efficacy at the end of physics 1 and physics 2 and their perceived recognition by TA/instructor in physics 2. Cohen suggested that typical values of d=0.2, 0.5 and 0.8 represent small, medium and large effect sizes, respectively [26]. Table I shows that female students had a higher average high school GPA than male students but lower average grades in both Physics 1 and 2 with small effect size. Table I also shows that there is a statistically significant gender gap in students' post self-efficacy scores both in physics 1 and physics 2 favoring male students with moderate effect sizes. In addition, Table I shows that female students also had lower perceived recognition by instructor/TA compared with male students with a moderate effect size.

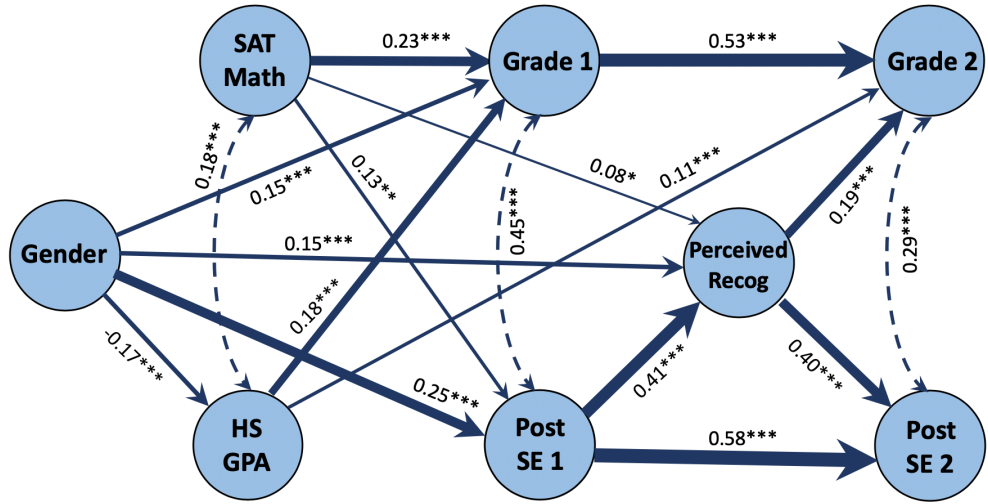

FIG 1. Results of the structural equation modeling with gender mediation showing **all** of the paths connecting various predictor and outcome variables for which the regression coefficients β are statistically significant (*p*-values are indicated by *** for *p*<0.001, ** for *p*<0.01 and * for *p*<0.05 values) [26]. Each arrow with the line connecting two variables in the diagram indicates the direction of regression. Numbers shown with regressions connecting two variables are standardized values of β that can be compared with each other and thicker lines qualitatively signify stronger β. Each dashed line with double arrow connecting two variables indicates the correlation between them, and the number on the line is standardized covariance between them.

Fig. 1 depicts the path analysis part of the SEM with gender mediation [26] for students' self-efficacy and grade at the end of physics 2 and how these are mediated by the perceived recognition by the course instructor/TA, controlling for students' high school math SAT scores, high school GPA as well as their grade and self-efficacy at the end of physics 1. While details of SEM will be presented elsewhere, our SEM model fits are CFI = 0.945, TLI = 0.925, RMSEA = 0.061 and SRMR = 0.035, which represent good fits [26].

In Fig. 1, the standardized regression coefficient β between each dependent (outcome) and independent (predictor) variable can be interpreted as the amount by which the dependent variable changes if the independent variable changes by one standard deviation [26]. We note that the β between perceived recognition by instructor/TA and student grade in physics 2 (Grade 2) is larger than the β between the high school GPA and Grade 2 suggesting that perceived recognition by their instructor/TA plays a more important role in predicting the course grade in physics 2 (see Fig. 1). Similarly, the impact of students' perceived recognition by instructor/TA on their post self-efficacy in physics 2 (β for Perceived Recognition to Grade 2) is large ( β = 0.40), and is roughly as large as the β = 0.58 between post self-efficacy in physics 1 and physics 2 (see Fig. 1). In addition, we note that gender does not directly predict students' grade and post self-efficacy in physics 2, even though there are statistically significant gender differences in students' grade and post self-efficacy in physics 2 as shown in Table I. This means that the gender differences at the end of physics 2 were actually mediated by students' perceived recognition in physics 2 in addition to students' grade and self-efficacy in physics 1. Thus, by recognizing students for their effort and progress when appropriate and being careful not to belittle them in unintended ways, instructors/TAs can impact their students' post self-efficacy and grade in physics 2 at the end of a two-term physics sequence in a positive way. This is especially true for students who already have lower self-efficacy due to societal stereotypes and biases about who can success in physics. In fact, instructor/TA's positive recognition for effort and progress at appropriate times can go a long way in supporting women's self-efficacy and identity and creating a learning environment in which all students can thrive [3, 20].

We note that prior investigation suggests that women with an A grade have comparable physics self-efficacy to men with a C grade in introductory physics [13] and this difference in self-efficacy has the potential to disproportionately drive out women from physics related majors and careers. In particular, the perception of not feeling positively recognized by the instructors/TAs as a person who can excel in physics has the potential to not only deteriorate students' self-efficacy and performance in a physics course but can also have a long term negative impact. While the details will be presented elsewhere, we note that even in individual interviews, we find that women are less likely than men to feel positively recognized by instructors/TAs (similar to Table I), and it has the potential to disrupt their paths away from physics related majors and/or careers. In the individual interviews we conducted with students, women sometimes reported that they had contemplated switching out of their majors (either engineering or physics) because of negative experiences in their physics courses while men did not express similar concerns, e.g., see Ref. [16]. The interviewed women sometimes noted that men in their physics courses were generally praised more by the instructor/TA than women and men often dominated asking questions and answering most of the questions instructors asked. Moreover, some women noted that instructors often did not make an effort to improve their sense of belonging, e.g., by ensuring that everyone was given an opportunity to contemplate the answers to the questions asked in small groups and then each time a different group was asked to answer the questions. Some interviewed women reported that sometimes instructors/TAs called men who answered the questions "brilliant", which made them feel as though they were not brilliant. They also noted that when they went to the course instructor or the TA to ask for help on physics problems, they were told that the problems were "easy", "obvious" or "trivial", which they perceived as disparaging or belittling in that they felt they were being told that they are not smart enough to do physics if they could not do such easy problems on their own. Some of them reported never going back to their instructor or TA after that for help.

**Discussion and Summary**: Women are severely underrepresented in physics and related disciplines in which there are pervasive gender stereotypes about who can excel and systemic disadvantages to women from these stereotypes accrue from a young age. We presented both quantitative and qualitative evidence focusing on the impact of male and female students' perceived recognition by the instructor and TA on their physics self-efficacy and performance at the end of a two-term calculus-based introductory physics sequence. We find gender differences in students' perceptions of being recognized.

Positively recognizing students who are underrepresented in physics is particularly important because lack of positive recognition has negatively impacted their entry and retention in physics related disciplines for decades. We strongly recommend that physics instructors and advisors have high expectations of all students. However, they should not be parsimonious in providing micro-affirmations to students whenever an opportunity arises and praise them for their effort instead of brilliance or intelligence [27]. Building an inclusive and welcoming environment, being empathetic, learning students' names as well as offering mentoring, guidance and support for next steps that focus on individual student's strengths, interests and growth can help in improving students' perception of feeling recognized [27]. Thus, instructors should set high standards but make it clear to students that they know they can achieve those high goals by working hard and working smart [28]. This is important for students to feel that the instructor was applying high standards to them and not just giving empty praise [28].

Also, calling problems that students are struggling with "trivial" or "obvious" is likely to have a student feel disparaged even if no harm was intended. What instructors and advisors must realize is that what is important is the impact on students of what they say and not whether harm was intended! Instructors should realize that labeling a question that a student is asking "trivial" has the potential to create stereotype threat [22,23] for students who are underrepresented in physics, increase their anxiety and decrease their physics self-efficacy.

We also encourage established members of the field to tackle head-on any issues of bias or microaggression that they observe in their classes rather than staying silent. Otherwise, due to societal stereotypes about physics, lack of perceived recognition as someone who can excel in physics has the potential to most hurt the underrepresented students, e.g., women who have few role models. They often enter physics courses with lower self-efficacy and doubts about whether they have what it takes to excel in physics courses. Our findings suggest that without explicit thought and action by instructors and TAs to appropriately recognize students as people who can excel in physics, not only is the gender gap in a physics course grade likely to be maintained but the self-efficacy gender gap is also likely to persist. This self-efficacy gender gap has the potential to not only have short term negative impact on women in physics courses but can also have long term negative impact, e.g., on their career choices.